\documentclass{scrartcl}
\usepackage{amsmath, amsfonts, amssymb}
\usepackage{physics}
\usepackage{natbib}
\usepackage{xcolor}
\usepackage{url}
\usepackage[utf8]{inputenc}

\allowdisplaybreaks
\usepackage[headsepline,footsepline]{scrlayer-scrpage}
\usepackage{bbm}
\usepackage{bm}

\usepackage{setspace}
\usepackage[varg]{txfonts}
\usepackage{etoolbox}
\usepackage{authblk}

\usepackage[colorlinks=false]{hyperref}
\usepackage{tocstyle}
\ihead{}  \ohead{}
\ifoot{}  \ofoot{}

\begin{document}

\title{A Brief Historical Perspective on the Consistent Histories Interpretation of Quantum Mechanics}
\author[1,4]{Gustavo Rodrigues Rocha}
\author[2,4]{Dean Rickles}
\author[3,4]{Florian J.\ Boge}

\affil[1]{\footnotesize Department of Physics, Universidade Estadual de Feira de Santana (UEFS),  Feira de Santana, Brazil}
\affil[2]{\footnotesize History and Philosophy of Science, University of Sydney, Sydney, Australia}
\affil[3]{\footnotesize Interdisciplinary Centre for Science and Technology Studies (IZWT), Wuppertal University, Wuppertal, Germany}
\affil[4]{\footnotesize Stellenbosch Institute for Advanced Study (STIAS), Wallenberg Research Centre at Stellenbosch University, Stellenbosch 7600, South Africa}

\date{}

\maketitle
\tableofcontents

\begin{abstract}
\noindent 
\end{abstract}

\section{Introduction}
Developed, with distinct flavors, between 1984 and the early 1990s, by Robert \cite{griffiths1984,griffiths1986}, Roland \cite{Omnes1987,Omnes1988a,Omnes1988b,Omnes1988c}, James \cite{Hartle1991,Hartle1993}, and Murray Gell-Mann (see for instance \cite{Gellmann-Hartle1990}), the consistent (or decoherent) histories approach to quantum mechanics, according to its proponents, aims at introducing ``probabilities into quantum mechanics in a fully consistent and physically meaningful way,'' which can be applied to a closed system (such as the entire universe), dispensing the notion of measurement (and, consequently, dissolving its inevitable paradoxes).

The formalism of the consistent/decoherent histories approach will be presented in \S\ref{formalism}, reviewing it in historical context in \S3. Robert Griffiths and Roland Omnès, whose seminal works will be explored in \S3 and \S4, considered their contributions to be ``a clarification of what is, by now, a standard approach to quantum probabilities'' \cite[321]{Freire2015}, namely, the Copenhagen perspective.

Gell-Mann and Hartle, on the other hand, who first applied the expression ``decoherence'' to histories \citep{Gellmann-Hartle1990,Gell_Mann-Hartle1994a,Gell_Mann-Hartle1994b}, following in the footsteps of \cite{zeh1970} Zeh (1970), \cite{Zurek1981}, and Joos (see for instance \cite{JoosZeh1985}), considered themselves to be post-Everettians, ascribing to Everett's 1957 work the merit of first suggesting ``how to generalize the Copenhagen framework so as to apply quantum mechanics to cosmology'' \cite[323]{Gell_Mann-Hartle1990b}.

Indeed, as it will be outlined in more details in \S\ref{sec:Dean}, Gell-Mann and Hartle had their sight set on quantum cosmology. However, instead of ``many worlds'' (Everett), they make use of the term ``many alternative histories of the universe'' \citep{Gell_Mann-Hartle1994a,Gell_Mann-Hartle1994b}, as ``branching'' (Everett) is replaced by ``decoherent histories'' in their interpretation. Gell-Mann and Hartle considered then decoherence history to be the right way to develop Everett's ideas.

Moreover, by introducing the role of ``information gathering and utilizing systems'' (IGUSes) in the formalism, Gell-Mann's and Hartle's emphasis on the applications of decoherent histories to cosmology and ``complex adaptive systems'' also added an ``evolutionary'' basis for our existence in the universe and for our preference for quasi-classical descriptions of nature \citep{Gell_Mann-Hartle1994b,Hartle1991}.

\section{The Formalism of the Consistent Histories Approach}\label{formalism}
\subsection{History Spaces}
The formalism of the consistent histories approach rests fundamentally on the notion of a \emph{history}, say $\alpha$, defined as a sequence of \emph{events}, $E_1,\ldots, E_n$, occurring, respectively, at times $t_1<\ldots<t_n$ \citep{griffiths1984}. In a classical setting, the $E_i$ could be, say, the falling of positions $\bm{x}_i$ of a particle performing a random walk into cells $V_j$ of a partition of some bounded subset of $\mathbb{R}^3$ at times $t_i$. The set of all sequences $\ev{\bm{x}_1\in V_j,\ldots,\bm{x}_n\in V_k}$ would then correspond to the set of possible histories, and at the same time define a \emph{sample space} \citep[110]{griffiths2003} or \emph{realm} (the quantum counterpart of \emph{sample space} in ordinary probability theory).

In quantum settings, events are represented by \emph{projectors} $\hat{P}_{k}$ that project to some (not necessarily one-dimensional) subspace of a suitable Hilbert space $\mathcal{H}$. To define an expression for the history of a system $S$ with Hilbert space $\mathcal{H}^{(S)}$, \cite{griffiths1996} introduces a \emph{different-time} tensor product, $\odot$, so that $\breve{\mathcal{H}}=\mathcal{H}^{(S)}_1\odot\ldots\odot\mathcal{H}^{(S)}_n$ corresponds to an $n$-fold tensor product of $S$'s Hilbert space considered at times $t_1,\ldots, t_n$. Accordingly, a history $\alpha$ can be represented by a \emph{history operator} $\hat{Y}_\alpha=\hat{P}_{1}\odot\ldots\odot\hat{P}_{n}$ on $\breve{\mathcal{H}}$, where each $\hat{P}_i$ corresponds to some projection on the respective $\mathcal{H}^{(S)}_i$ (i.e., a projection on $\mathcal{H}^{(S)}$ considered at time $t_i$).

Defining histories in this way, one can also introduce logical operations on them \citep[e.g.][113 ff.]{griffiths2003}. However, this is straightforward only in case $\mathcal{H}^{(S)}$ is finite-dimensional \citep[see][\S VIII B]{griffiths1996}. The intricacies of extending the formalism to infinite-dimensional spaces have been explored, e.g., by \cite{Isham1, Isham2}, but shall not be covered here. 

In the finite-dimensional case, the (negated) proposition `$\alpha$ did not occur', for instance, can be defined by $\hat{Y}_\alpha^\perp:=\breve{\mathbbm{1}}-\hat{Y}_\alpha$, where $\breve{\mathbbm{1}}$ is the identity on $\breve{\mathcal{H}}$. Defining pendants for logical operations in this way, it is then possible to associate probabilities to histories, and to regard  as a \emph{sample space} any set of histories with operators $\hat{Y}_\alpha$ that satisfy $\sum_\alpha\hat{Y}_\alpha= \breve{\mathbbm{1}}, \hat{Y}_\alpha\hat{Y}_\beta=\delta_{\alpha\beta}\hat{Y}_\alpha$ i.e., where any two distinct histories (or rather: their operators) are mutually orthogonal, and where they jointly exhaust the entire space (their operators sum to the unity). 

In quantum theory, those probabilities are, of course, given by the Born rule. Application to histories is only possible, however, if the dynamics in between events are specified. This is  most conveniently established by defining so-called \emph{chain operators}. A chain operator $\hat{C}_{\alpha}$ for history $\alpha$ with operator $\hat{Y}_\alpha=\hat{P}_{1}\odot\ldots\odot\hat{P}_{n}$ on $\breve{\mathcal{H}}$ is given by
\begin{equation}
\hat{C}_{\alpha}:= \hat{P}_{n}\hat{U}(t_n;t_{n-1})\hat{P}_{n-1}\hat{U}(t_{n-1};t_{n-2})\ldots\hat{U}(t_{3};t_{2})\hat{P}_{2}\hat{U}(t_2;t_{1})\hat{P}_{1},
\end{equation} 
where the $\hat{U}(t_i; t_{i-1})$ are unitary operators representing the time-evolution between events. For instance, if the dynamics are governed by a time-independent Hamiltonian $\hat{H}$, then $\hat{U}(t_i; t_{i-1})=e^{\frac{\imath}{\hbar}\hat{H}(t_{i-1}-t_{i})}$. It should be clear that the step from $\hat{Y}_\alpha$  to $\hat{C}_{\alpha}$ is thus non-trivial: different choices of dynamics will result in different chain operators. 

It is useful to  also introduce a second index $a$ on the projectors, so they be distinguished not only by their application at points in time but also by their function to represent different events. For instance, $a_i$ could be the value an observable $A$ takes on at $t_i$, so that $\sum_{a_i}\hat{P}_{a_i}=\mathbbm{1}$ ($\mathbbm{1}$ the identity on one copy of the space $\mathcal{H}^{(S)}$). We here use $a$ as a generic label, however, which need not be associated with the same observable $A$ in each occurrence. Using this notation, one may also define a Heisenberg operator
\begin{equation}
\hat{P}_{a_i}^{i}:=e^{\frac{\imath}{\hbar}\hat{H}t_{i}}\hat{P}_{a_i}e^{-\frac{\imath}{\hbar}\hat{H}t_{i}},
\end{equation}
so that $\hat{C}_{\alpha}$ reduces to
\begin{equation}
\hat{C}_{\alpha}:=\hat{P}^{n}_{a_n}\hat{P}^{n-1}_{a_{n-1}}\ldots\hat{P}^{1}_{a_1}.
\end{equation} 
Using chain operators, we retrieve the expression 
\begin{equation}\label{eq:histpro}
p(\alpha) = p(\wedge_i E_i) = \Tr[\hat{C}_{\alpha}^\dagger\hat{C}_{\alpha}]
\end{equation} 
for the probability of $\alpha$'s occurrence, where Tr denotes the trace. Moreover, if a (normalised) density operator $\hat{\rho}_S$ can be associated to $S$ at some initial time, we can extend this to 

\begin{equation}\label{eq:Prob}
p(\alpha|\hat{\rho}_S) = \Tr[\hat{C}_{\alpha}^\dagger\hat{\rho}_S\hat{C}_{\alpha}]
\end{equation}
\citep[5455]{Isham2}.

\subsection{Consistency, Compatibility, and Grain}
Understanding Griffiths' \emph{consistency conditions} for histories requires to think of the trace functional as a more general `weight', $w$, not necessarily to be identified with a probability under all circumstances. If the weight $w(\alpha\wedge\beta)=\Tr[\hat{C}_{\alpha}^\dagger\hat{C}_{\beta}]$ jointly assigned to histories $\alpha\neq\beta$ \emph{vanishes}, history operators $\hat{Y}_\alpha, \hat{Y}_\beta$ satisfy a pendant of an additivity condition $p(A\vee B)=p(A)+ p(B)$ for $A,B$ mutually exclusive events. Under these conditions, $\alpha$ and $\beta$ may be called \emph{consistent}. A more straightforward way of putting this is ``that the chain operators associated with the different histories in the sample space be mutually orthogonal in terms of the inner product defined [by $\Tr[\hat{C}_{\alpha}^\dagger\hat{C}_{\beta}]$]'' \citep[141]{griffiths2003}.\footnote{That this indeed defines an inner product can be seen by investigating its properties \citep[positive definiteness, skew-symmetry, and anti-linearity in one argument; see][139]{griffiths2003}.}

Additionally, one may introduce the notion of a \emph{family of histories}, $\mathcal{F}$, defined as a set of sequences $\alpha=\ev*{E_{1},\ldots,E_{n}}, \beta=\ev*{\tilde{E}_{1},\ldots,\tilde{E}_{n}}, \ldots$, such that at each point of any of the sequences that constitute the family, one out of a set of mutually exclusive and jointly exhaustive events occurs. In other words: that the event set defines a sample space of events possible at that point in time. A family $\mathcal{F}$ of histories is said to be \emph{consistent}, moreover, if any two $\alpha\neq\beta$ from $\mathcal{F}$ are, and both a consistent family and its operator-pendant are also called \emph{framework} by \citet[141]{griffiths2003}. The centerpiece of Griffiths' work is the suggestion that ``only those sequences, or histories, satisfying a consistency requirement are considered meaningful'' \citep[516]{griffiths1986}. 


An equally important but weaker notion is that of \emph{compatibility}, which will be further explored in \S\ref{sec:Gustavo}. Two families, $\mathcal{F}$ and $\mathcal{G}$, are said to be compatible just in case they have a common \emph{refinement} that is itself consistent \citep[147]{griffiths2003}, and otherwise are called \emph{mutually incompatible}. `Refinement' here refers to the process of resolving a projector $\hat{P}$ in terms of projectors $\hat{P}', \hat{P}''$ that project onto subspaces of the space projected onto by $\hat{P}$. Conversely, if we sum $\hat{P}', \hat{P}''$ to get $\hat{P}$, we may call this a \emph{coarsening} (or coarse-graining) of the event space.

\subsection{Decoherence}
As a final point, consider the connection to \emph{decoherence}, made especially by \cite{Gell_Mann-Hartle1990b}. Decoherence, as originally discovered by \citet{zeh1970}, refers to the dynamic vanishing of interference due to the interaction of a system with its environment. In brief, if $S$ couples to environment $E$ in a suitable way, and the dynamics is dominated by the interaction, it is possible to show that the effective (pure state) density matrix
\begin{equation}
 \hat{\rho}_{S}=\Tr_{E}[\hat{\rho}_{SE}] = \sum_{i,j} \alpha_{i}\alpha_{j}^{*}\dyad*{\mathcal{S}_{i}}{\mathcal{S}_{j}}\Tr[\dyad*{\mathcal{E}_{i}}{\mathcal{E}_{j}}] =\ldots = \sum_{i,j} \alpha_{i}\alpha_{j}^{*}\dyad*{\mathcal{S}_{i}}{\mathcal{S}_{j}}\braket*{\mathcal{E}_{j}}{\mathcal{E}_{i}},
\end{equation}
becomes $\hat{\rho}_{S} \approx \sum_{j} |\alpha_{j}|^{2}\dyad*{\mathcal{S}_{i}}{\mathcal{S}_{j}}$, because $\braket*{\mathcal{E}_{j}}{\mathcal{E}_{i}}$ vanishes dynamically for $i\neq j$.

The problems with interpreting decoherence directly are well known: It does not lead to an `or' from an `and' \citep{bell1990}, and the possibility of recoherence does not favour a straightforward epistemic reading either, even though $\hat{\rho}_{S}$ looks suspiciously like a statistical mixture. In any case, many-worlds theorists have counted decoherence in their favour for reasons touched on below.

As far as the consistent histories apporach is concerend, \cite{Gell_Mann-Hartle1990b} realised, apparently independently of Griffiths and Omn\`es, that the expression $ \Tr[\hat{C}_{\alpha}^\dagger\hat{\rho}_0\hat{C}_{\alpha}]$, with $\hat{\rho}_0$ ``the initial density matrix of the universe'' (ibid., 326), corresponds to a complex functional $D(\alpha, \beta)$ that can be interpreted in terms of the decoherence between histories $\alpha, \beta$.  

The reference to the initial state of the universe here indicates Gell-Mann and Hartle's motivation of providing a foundation for quantum cosmology. However, decoherence in the sense of Zeh and others originally meant the vanishing of coherence due to the fact that essentially any system is `open'. Considerations of decoherence for the sake of cosmology not just require ``assuming[...] that the key features [...][of decoherence] transcend the Hilbert-space language'' \citep[2915--6]{halliwell1989}, but to think of the universe as one \emph{closed system}, whose internal dynamics provides consistency conditions for situated, reasoning agents (see \S\ref{sec:Dean} for details). 

Intuitively, when $D$ vanishes for two histories $\alpha\neq\beta$, they have `nothing in common' and thus do not interfere. This coincides with the notion of consistency adopted, at least since 1994, by Griffiths (see \cite{griffiths1994}). However, it is, first, important to note that this can only happen trivially for two \emph{completely fine-grained} histories, i.e., where every projector is one-dimensional.

The sense of triviality here is that ``what happens at the last moment is uniquely correlated with the alternatives at all previous moments'' \citep[13]{Gell_Mann-Hartle2007}. Hence, histories are always distinguished and decoherence does not do any real work. Using that $D(\alpha, \beta)$ reduces to $\ev{\hat{C}_\alpha^\dagger\hat{C}_\beta}{\Psi}$ for a pure initial state. It can be shown that there is only trivial decoherence of completely fine-grained histories in this sense.

Now, second, for \emph{coarse-grained} histories $\alpha, \beta$, where $D(\alpha, \beta)$ can also vanish non-trivially, one can distinguish at least two `levels' of decoherence: 
\begin{align}
D(\alpha, \beta) &=p(\alpha)\delta_{\alpha\beta} \tag{\text{medium decoherence}}\\
\text{Re}[D(\alpha, \beta)] &=p(\alpha)\delta_{\alpha\beta} \tag{\text{weak decoherence}}.
\end{align}

A good motivation for seeing that the weaker condition also quantifies decoherence is that, if $\bar{\alpha}$ is a coarsening of $\alpha$ and $\beta$, in the sense that $\hat{C}_{\bar{\alpha}} =\hat{C}_\alpha + \hat{C}_\beta $, and $\hat{\rho}_0=\dyad*{\Psi}$, we have
\begin{equation}
p(\bar{\alpha}|\hat{\rho}_S) = \ev{\hat{C}_{\bar{\alpha}}^\dagger\hat{C}_{\bar{\beta}}}{\Psi} = \ldots =p(\alpha) + p(\beta) + 2\text{Re}\ev{\hat{C}_\alpha^\dagger\hat{C}_\beta}{\Psi}, 
\end{equation}
which only gives back the classical sum over unresolved possibilities in case the final term vanishes. $\hat{C}_{\bar{\alpha}}$ here is not necessarily itself a chain of time-dependent projectors, and thus called a \emph{class operator} \citep[cf.][]{hartle1995}. Furthermore, a family of histories of which any two are at least weakly decoherent is also called a \emph{realm} by \citet{Gellmann-Hartle1990}, which notion thus delivers a generalization of Griffiths' notion of a framework. 

\citet[242]{jooszehkiefer2003} have actually lamented that, in contrast to weak decoherence, medium decoherence 
\begin{quote}
can also be fulfilled for very non-classical situations [...], [so] we prefer to call this a consistency condition to avoid confusion with the physical concept of decoherence [...].
\end{quote}

Nevertheless, Gell-Mann and Hartle have investigated even stronger notions of decoherence. In \cite{Gell_Mann-Hartle1994a}, an elaborate construction is presented, which, among other things, allowed them to establish a ``connection [...] between the notion of strong decoherence and the work of many authors who investigate the evolution of a reduced density matrix for a system in the presence of an environment'' (ibid., 24) or that ``strong decoherence implies the diagonalization of a variety of reduced density matrices''  (ibid., 26). 

\section{Interpreting and Understanding the Formalism in Historical Context}\label{sec:Gustavo}
\subsection{Single-Framework Rule and Realm-Dependent Reality: From Method to Reality and Back}
As described in \S\ref{formalism} above, a \emph{history} is a sequence of events, $E_1,\dots, E_n$, represented by \emph{projectors} $\hat{P}_{k}$ on the Hilbert space $\mathcal{H}$ at times $t_1 <\ldots< t_n$. Moreover, we have seen that it is only allowed to have probabilities assigned to a \emph{family} (or set) of histories, $\mathcal{F}$, providing it fulfills Griffith's consistency conditions, namely, (i) the sum of probabilities of the histories in the family equals one, and (ii) all pairs of histories within the family are orthogonal.

A family of histories that meets these conditions is called a \emph{consistent framework} (or just \emph{framework} for short) or \emph{realm }(which constitute, as seen in \S\ref{formalism}, the quantum counterpart of \emph{sample spaces} in ordinary probability theory). In other words, a \emph{realm} is a decoherent set of alternative coarse-grained histories or, as Griffiths would phrase it, ``a \emph{framework} is a Boolean algebra of commuting projectors based upon a suitable sample space'' \citep[216]{griffiths2003}. Finally, we have seen that \emph{different realms can be mutually incompatible}.

Thus, an important feature of the consistent histories approach is this notion, suggesting a \emph{realm-dependent reality}, that quantum systems possess \emph{multiple incompatible frameworks}. Hartle claims that the reason we hardly realize why this is the only way to make meaningful statements is because human languages contain ``excess baggage'' \citep{Hartle2007}.

Indeed, the so-called \emph{single-framework rule} (which states that two or more incompatible quantum descriptions cannot be combined to form a meaningful quantum description) must be followed or enforced in order to avoid inconsistencies \citep{Griffiths2009}. ``Once this notion has been accepted (or ‘digested')'', as P. C. Hohenberg put it well, ``the theory unfolds in a logical manner without further paradoxes or mysteries'' \citep[2843]{hohenberg2010}.

A historian of science might then wonder within what sort of historical context these premises would be more prone to be accepted (or ``digested''). What does it take for one to accept the idea (following from the formalism) of a \emph{realm-dependent reality}?
In what follows, it will be suggested that this context is twofold: i) that of an informational worldview based on previous attempts (going as far back as to the turn of the 20th century) to formalize the sciences (what will be seen in §3, followed by additional information about Griffiths’ and Omnès’ trajectories in §4) and ii) that of having the universe as a whole taken as an object of science (the ultimate closed system, as it will be seen in §5, based on the works by Gell-Mann and Hartle).

To begin with the former (the importance of formalization to the epistemology behind the consistent histories interpretation), an exposition of Omnès’ thoughts will be outlined in the following subsections.

Furthermore, as it will be seen in §3.3 and §5, as these two contexts come together in the last quarter of the 20th century, the importance of the study of complex adaptive systems becomes more relevant and apparent to the consistent histories approach, as epitomized by the introduction of IGUSes in the formalism.

The main argument in the following two subsections will be that the concept of a realm-dependent reality seems to rely on a sort of “structural objectivity” (see \citeauthor{DastonGalison2007}, \citeyear{DastonGalison2007}), as formalism should be taken as more trustful than experience and common sense in the making of our metaphysics and epistemology.

Omnès states that very clearly about the method one should follow to interpret quantum theory: “we must not begin with the classical but with the quantum world, and deduce the former together with all its appearances” \citep[164]{Omnes1994}, i.e., “to deduce common sense from the quantum premises,” as best formulated by a formal logic/ a formal physics – being formalism the utmost reach of modern science, as if we had achieved a certain historical “stage of reason” \citep[191]{Foucault1969}.

Indeed, Omnès concludes that our current mature “scientific reasoning” follows (or should follow) a four-stage method, i.e., “empiricism, concept formation, development and verification” (epitomized by the history of science itself), by which “reality is summoned twice, at the beginning” (a more empiricist phase) and at the end of the process (when formalism gradually imposes itself) (i.e., from method to reality and back). Omnès’ reference to the history of science encapsulates his worldview.

\subsection{Information, Formalization, and the Emergence of Structural Objectivity}

Behind the acceptance of an informational worldview underlies a generalized practice of formalism, which could be seen, following Hacking, as a \emph{style of reasoning}, “a matter not only of thought but of action,” constituted by “techniques of thinking” \citep{Hacking1990}. It is no accident that Omnès calls our time not as the “Information Age,” but as the “Age of Formalism” \citep{Omnes1994}. The “Information Age” was in certain sense the end result of the “Age of Formalism,” since, historically, the later came first (paving the way to the former) and eventually became embodied in the technological advances of the second half of the 20th century.

In fact, just as \cite{Hacking1990} claimed that it took from around 1660 to 1900 for us to move from two different \emph{style of reasoning}, the former characterized by universal Cartesian causation and determinism, as central concepts both of science and nature (at a time when chance was seen as superstition, vulgarity and unreason) to the latter, a “universe of absolute irreducible chance” (where “nature is at bottom irreducibly stochastic”), Omnès also claimed that our path to the \emph{style of reasoning} of formalism took us “approximately one hundred years, from 1850 to 1950. Several names stand out: Weierstrass, Dedekind, Cantor, Frege, Peano, Hilbert, Russel, and Whitehead” \citep[102]{Omnes1994}.

The concept of “information” became in the fourth quarter of the twentieth century what cultural historian Arthur Lovejoy used to call a \emph{unit-idea}, a primary and recurrent dynamic unit which blends an aggregate of “implicit or incompletely explicit assumptions, or more or less unconscious mental habits, operating in the thought of a generation” \citep[7]{Lovejoy1936} or what the philosopher Michael Foucault once called an \emph{episteme}, “a worldview, a slice of history common to all branches of knowledge, which imposes on each one the same norms and postulates, a general stage of reason, a certain structure of thought that the men of a particular period cannot escape” \citep[191]{Foucault1969}.

The interplay between metaphysics and physical sciences were well explored by historians and philosophers of science such as Pierre \cite{Duhem1906}, Emile \cite{Meyerson1098}, and E. A. \cite{Burtt1924} – who somehow disclosed the metaphysics underlying our physical theories. The concept of “information”, from the late 1970s onwards, started playing the role of our constitutive metaphysics, just as the concepts of “matter”, “force” and “energy”, from the seventeenth to the nineteenth century. Meanwhile, inasmuch as “information” was being consolidated as \emph{unit-idea}, quantum field theory was becoming our most basic theory to describe the physical world.

This historically constructed informational metaphysics eventually interwove with our foremost physical theories. A most revealing text in this regard is Omnès’ “Quantum Philosophy: Understanding and Interpreting Contemporary Science.” As we have seen, formalization is a key idea behind the central concept of information underlying our most fundamental metaphysics. Omnès is aware that the founding fathers of quantum mechanics, such as Niels Bohr, did not have this powerful informational/ computational metaphor at hand when formulating the puzzles emerging from quantum theory: “many of these conclusions, by the way, had already been guessed by Bohr thanks to a stroke of genius, unaided by the powerful mathematical techniques and the guide of a discursive construction available to latecomers” \citep[233]{Omnes1994}.

\subsection{Cybernetics and Measurement: From von Neumann and the Cybernetics Group to the Santa Fe Institute}
It has been pointed out that one important feature of the consistent histories approach is that it does not depend on an assumed separation of classical and quantum domain, on notions of measurement, or on collapse of the wave function. Indeed, there is no role assigned to observers (or measurement) in consistent histories. How does the project of the formalization of the sciences relates to these features?

In setting an observer-independent criterion, consistent histories approach was following a historical trend (well noticed by Omnès) that had already been sweeping all branches of knowledge by the last quarter of the 20th century, i.e., structural objectivity – an radicalization in certain sense of a former mechanical objectivity (based on the idea of representation and the ideal of truth as correspondence to nature).

Both mechanical and structural objectivity shared a common enemy, i.e., subjectivity.  However, structural objectivity, as opposed to mechanical objectivity, gave up “one’s own sensations and ideas in favor of formal structures accessible to all thinking beings” \citep[257]{DastonGalison2007}. Moreover, structural objectivity is an even stronger instantiation of a trend for objectivation in modern science: “Structural objectivity was, in some senses, an intensification of mechanical objectivity, more royalist than the king” (ibid, 259).

If mechanical objectivity “had been a solitary and paradoxically egotistical pursuit: the restraint of the self by the self,” as instantiated by the transcendental subject in Kantianism, structural objectivity, in contrast, “demanded self-effacement” (ibid,, 300) or, to rephrase it, an even more radical desubjectivation. Although a plethora of names and research programs across a wide range of disciplines could be mentioned, cybernetics and its offspring (e.g., system theories) will be the focus of this section.

A number of authors \citep{Heims1991,Dupy1994,Lafontaine2004} have emphasized the seminal importance of the Cybernetics Group as a crucial move towards this direction.

A series of multidisciplinary conferences, supported by the Macy Foundation, was held, between 1946 and 1953, to discuss a variety of topics that came to be called as cybernetics. Members of the so-called Cybernetics Group included mathematicians John von Neumann and Norbert Wiener, anthropologists Margaret Mead and Gregory Bateson, psychologist Kurt Lewin, neuropsychiatrist Warren McCulloch, among many others.

Lafontaine has convincedly argued that “there is a certain paradigmatic unit underlying this imponent theorical diversity: from structuralism to systems theory, from postmodernism to posthumanism, from cyberspace to biotechnological reshaping of the bodies, in any case, it can be verified, over and over again, the same denial of the humanist heritage, the same logic of desubjectivation” \citep[17]{Lafontaine2004}.

It sounds as one of Lovejoy’s \emph{unit-idea} or Foucault’s \emph{episteme}. Dupuy has also called the attention to the fact that although “cybernetics was soon forgotten, apparently consigned to a dark corner of modern intellectual history,” several decades later, it reemerged, and, after having undergone a mutation, now bears “the features of the various disciplines that make up what today is known as cognitive science” \citep[ix]{Dupy1994}.

The formalization of the mind, processes and behaviors, the coadunation of the understanding of animals, humans and machines, led to a “subjectless philosophy of mind.” As pinpointed by Dupuy: “consider perhaps the most significant contribution of cybernetics to philosophy: mind minus the subject” \citep[107]{Dupy1994}.

The informational metaphor was of fundamental importance for this \emph{Weltanschauung}. The Cybernetics Group’s research program was inspired and informed by behaviorism, but it took its functionalism to the next level. The replacement of the concept of “behavior” by the concept of “information” had enormous consequences, as these models can be indistinguishably applied to living systems and machines.

In the early 1950s, for instance, the father of information theory, Claude Shannon, who had studied under Norbert Wiener at MIT, demonstrated how “Theseus,” a life-sized magnetic mouse controlled by relay circuits (a perfect correlate to the paradigmatic mice experiments of behaviorism),  could learn his way around a maze. 

If the concept of “behavior” didn’t distinguish humans from animals, the concept of “information” pushes it even further, as it doesn’t differentiate the living from the non-living, the subject being dissolved into a physical symbol system. It is no wonder that, within this context, as the mouse and the maze form one single systemic network of information exchange, it is nonsensical to separate them substantially.

As an incipient example of a complex adaptive system, Theseus is in structural coupling with the maze, adapting itself through (positive and negative) feedback loops to its environment (in the language of cybernetics this would be called a “servomechanism”). The observer-independent criterion of consistent histories approach, summarized by the single-framework rule, likewise, follows a similar cybernetic philosophy of nature.

As cognition in cybernetics and system theories is a process (and not a “thing,” such as “mind”), extracting information about a complex adaptive system has nothing to do with an act of measurement in the usual sense of an observer and an observed. It is rather a quest of selecting a meaningful framework in an indefinite network of interactions. It is also the result of applying formal operations that, although drawn from the empirical inquiry of nature, cannot be reducible to any of our common intuitions.

Drawing the consequences of applying this \emph{style of reasoning} to the consistent histories approach seems to be straightforward:

i) the need to divide the world into a system and an external observer becomes inoperative and ii) the concept of history gains a fresh meaning as the evolution of a complex adaptive system.

Jeffrey Barrett and Peter Byrne, in their comments on the letters exchange between Everett and Wheeler, as well as Everett and Norbert Wiener, have rightly pointed out a generation change taking place in interpreting the paradoxes of quantum mechanics:

“Information theory was a starting point for Everett, and it is not surprising that physicists of Bohr’s and Stern’s generation tended to think of information in terms of ‘meaning,’ whereas Everett thought of information as a formal notion that might be represented in the state of almost any physical system ¬– in keeping with his background in game theory and the new science of ‘cybernetics.’ That is perhaps why Everett could easily conceive of an observer as a servomechanism, whereas Bohr (a neo-Kantian) and Stern (a Bohrian) could not separate measurement from human agency” \citep[214]{Barret-Byrne2012}.

For someone who embraces formalism as the foremost realization of the progress of modern science, the notion, in consistent quantum approach, that reality, or what is real, is meaningful only relative to a \emph{framework} should sound just as natural as the notion, in special relativity, that simultaneity is relative to a reference frame. Both concepts, although counter-intuitive, are products of formalism (having Einstein provided an interpretation to Lorentz transformation). As Omnès would agree:

“With the advent of relativity, the theory of knowledge has forever ceased to be cast in an intuitive representation, to be based solely on concepts whose only authentic formulation involves a mathematical formalism” \citep[197]{Omnes1994}.

In the same manner, for Omnès, the interpretation of quantum mechanics should follow from a small set of basic principles, not a host of ad hoc rules needed to deal with particular cases, as it should follow from the formalism to the common sense of experience (and not the other way around).

Omnès, indeed, brought to our attention, in his 1994 book, how the consistent histories approach mirrors and departs from Kant’s philosophy which Omnès has in high respect. However, he seems to have missed an important feature in his comparison.

Omnès summarized the measurement problem and the function of interpretation as following: “the alpha of facts seems to contradict the omega of theory” and “the purpose of interpretation is to reconcile these extremes,” and found in Kant’s unsolvable antinomies, the twelve categories of understanding and the forms of sensible intuition (i.e., space and time) an unparalleled attempt of formalization of physical sciences. However, Omnès opposes to the \emph{synthetic a priori} character of Kant’s proposal

“the roots of logic, if not of reason, are not to be sought within the structure of our mind but outside it, in physical reality. This of course implies a complete reversal of Kant’s approach” \citep[76]{Omnes1994}. For Omnès, even though not quite as Kant had in mind, formalism would have taken Kant’s proposal to fruition in the twentieth century.

However, Omnès’ narrative missed an important aspect of Kantianism. Back in 1963, Ricoeur, when commenting about the structuralism of Claude Lévi-Strauss and its resemblances with Kantianism, got it right to the point: “Kantianism, yes, but without the transcendental subject” \citep[618]{Ricoeur1963}. As a matter of fact, the formalization of all branches of scientific knowledge culminates in the end of a long modern history of the “metaphysics of subjectivity” – to use Heidegger’s expression (whose hostility towards cybernetics was well known at the period).

Similarly, by means of the use of appropriate frameworks, the separation of the system from the measurement apparatus becomes pointless in the consistent history approach – as “measurement” is inextricably subsumed in what Griffiths calls “quantum reasoning,” which is nothing but the usual rules for manipulating probabilities providing the \emph{single-framework rule} is enforced.

Would that quantum reasoning be so convincingly in a different historical context (other than the Age of Formalism)? As Hacking suggested, \emph{styles of reasoning} can be sometimes incommensurable in comparison with backgrounds so different in their historical contexts.

Lafontaine devotes two chapters in her monograph, “Colonization 1: The Structural Subject” and “Colonization 2: The Systemic Subject,” to document the “colonization” of cybernetics over, respectively, structuralism and system theories/ complex theories, the latter being part of the so-called second cybernetics. Dupuy mentions three offspring coming out of the Cybernetics Group:

“one of these, an offspring of the second cybernetics, grew out of the various attempts made during the 1970s and 1980s to formalize self-organizing in biological systems using Boolean automata networks; in France this work was conducted chiefly by a team directed by Henri Atlan and including Françoise Fogelman-Soulié, Gérard Weisbuch, and Maurice Milgran; in the United States, by Stuart Kauffman, a former student of Warren McCulloch, and others at the Santa Fe Institute for the Study of Complex Systems. Another, also springing from the second cybernetic, was the Chilean school of autopoiesis” \citep[103]{Dupy1994}.

The Santa Fe Institute (SFI) began functioning as an independent institution in 1984 and included among its co-founders George Cowan, David Pines, and Murray Gell-Mann \citep{Pines1988}. “The Santa Fe Institute,” described Gell-Mann, “is devoted to the study of complex systems, including their relation to the simple laws that underlies them, but emphasizing the behavior of the complex systems themselves” \citep{Gell_Mann1990}.

Stuart Kauffman, a former student of Warren McCulloch, the key figure of the Cybernetics Group, approached Cowan and joined the SFI. As a multidisciplinary research and teaching institute, remembered Cowan, “we struggled to find a term that would embrace their commonalities and settled on ‘Complexity.’ Later our mantra became ‘Complex Adaptive Systems’” \citep[144]{Cowan2010}.

As it will be seen in §5, concepts such as IGUSes (“information gathering and utilizing systems”), developed by Gell-Mann and Hartle in their consistent histories approach, clearly convey the episteme (in the sense given by Foucault) underlying the Santa Fe Institute.

\section{Robert Griffiths and Roland Omn\`{e}s: Context and Logic}\label{sec:GriffOm}
Having reflected upon the philosophical implications drawn from consistent histories by Omnès in \S3, in the following some extra important facts will be given about Omnès and Griffiths trajectory in formulating the consistent histories approach.

Robert Griffiths is a statistical physicist at Carnegie Mellon University, having studied Physics in Princeton and at Stanford University. After a productive career on statistical mechanics in the 1960s and 1970s, Griffiths turned his attention to foundational issues of quantum mechanics in the 1980s, working on the consistent histories approach thereafter, and publishing his major work on the subject in 2002.

When Griffiths first introduced the concept of “consistent histories”, as opposed to “measurement”, in his 1984 paper, “Consistent Histories and Interpretation of Quantum Mechanics,” the idea of “history” had already been put forward, although with different meaning, by R. P. Feynman and P. A. M. Dirac. It is meant by “consistent” that Griffiths’ proposal contains no paradoxes – as it can be used in the study of closed systems (not requiring the artificial separation between the classical and quantum domains and without the need to divide the world into a system and an observer).

In a follow-up 1986 paper, “Making Consistent Inferences from Quantum Measurements,” Griffiths made it clearer as for to what motivated him to seek for a way to reasoning consistently in quantum settings:

\begin{quote}It is widely acknowledged that there are conceptual difficulties (in particular, problems of logical consistency) in making inferences about microscopic phenomena from the results of experimental measurements. [...] It seems to me that it is possible to speak sensibly about microscopic phenomena and how they are related to measurements. However, the procedure for doing this, namely, the logical process of inference, must be appropriate to the domain of study and must be consistent with and indeed based on quantum theory itself \citep[512]{griffiths1986}. \end{quote}

“Following essentially Griffiths’ proposal”, Omnès, afterwards, noticing the logical background behind the consistent histories, constructed a “consistent Boolean logics describing the history of a system” (Omnès, 1988a, 893). Omnès claimed that to be a progress over Griffiths’ proposal. Both proponents of the consistent histories approach, in the end, aimed at the same, i.e., a consistent and complete reformulation of Bohr’s view. Niels Bohr presented the complementary principle, as the essence of quantum mechanics, at the Solvay Conference in 1927. Having been supported by a number of physicists this view became known as the “Copenhagen Interpretation”. 

In fact, in his 1994 book, Omnès reminded his readers that consistent histories “might be considered as being simply a modernized version of the interpretation first proposed by Bohr in the early days of quantum mechanics” \citep[498]{Omnes1994} and Griffiths, in the same vein, claimed that the consistent histories approach could be seen as merely consisting of “Copenhagen done right” \citep[2835]{hohenberg2010} – although, it should be mentioned, historians and philosophers alike often point out that there is no such thing as a monolithic view of the Copenhagen interpretation.

Roland Omnès is an Emeritus Professor at the University of Paris XI in Orsay, having begun to work on an axiomatization of quantum mechanics in his 1987 paper, ``Interpretation of Quantum Mechanics.'' As one can read in his final note by the end of his paper, when asked by a referee how his approach differed from Griffiths' consistent histories, Omnès replied that, ``as far as mathematical techniques are concerned, Griffith's construction is used,'' however, ``the conceptual foundations are different because what is proposed is a revision of the logical foundations of quantum mechanics'' \citep[172]{Omnes1987}.

As a matter of fact, in a series of papers in the following couple of years, titled “Logical Reformulation of Quantum Mechanics,” Omnès fully developed this idea \citep{Omnes1988a,Omnes1988b,Omnes1988c,Omnes1989}. In the mid 1990s, Omnès developed even more completely the formalism, besides his reflections on historical, epistemological and philosophical aspects of the consistent histories approach in three major textbooks \citep{Omnes1994,Omnes1999,Omnes-Sangalli1999}, having his last word so far on the philosophical aspects of physics and mathematics been expressed more recently in a 2005 book \citep{Omnes2005}.

\section{Decoherent Histories: Consistent Histories \`{A} La Gell-Mann and Hartle}\label{sec:Dean}
Murray Gell-Mann (1929-2019) was the quintessential particle theorist. James Hartle (born 1939), a product (as a postdoc as well as an undergraduate student) of John Wheeler's Princeton relativity school, is the  quintessential cosmologist, though he also possesses strong particle physics acumen, having  been a PhD student of Gell-Mann's in fact. Together, beginning in the late 1980s (see Gell-Mann and Hartle, 1994a), they introduced the so-called ``decoherent histories'' approach to quantum mechanics, which can be viewed largely as an application of the consistent histories approach of Griffiths and Omn\`{e}s to cosmology, as described by classical general relativity, with the lessons of Everett and decoherence thrown into the mix.

Though apparently discovered independently of these earlier approaches, Gell-Mann and Hartle do cite Griffiths' and Omn\`{e}s' works in their initial paper, noting  at least consistency with them, despite their own wider perspective. Whereas the consistent histories approaches of Griffiths and Omn\`{e}s have their roots in making sense of any sequence of events in a subsystem, with every element described solely quantum mechanically (as described in \S\ref{formalism} above), the approach of Gell-Mann and Hartle focuses on  histories in a more global sense corresponding to the possible histories of the universe understood as a system in its own right: the ultimate closed system.  Griffiths and Omn\`{e}s had no such grand aspirations.  Expanding consistent histories to the universe brings in a host of additional conceptual and technical problems, especially involving the emergence of a seemingly classical world with space, time, and observers experiencing an apparently unique history. They sought a framework, as Hartle puts it, "within which to erect a fundamental description of the universe which would encompass all scales -- from the microscopic scales of the elementary particle interactions to the most distant reaches of the realm of the galaxies - from the moment of the big bang to the most distant future that one can contemplate" \citep[1]{Hartle1992}.

Obviously, to a quantum cosmologist, an approach which does minimal violence to quantum mechanics yet allows it to be applied to genuinely closed systems (with no reference to measurement or observers as having special importance) is almost a holy grail. Of course, there had already been the Everett interpretation which made the same claims, but that seemed either too extravagant or unable to explain its basic process of branching (and the related emergence of a classical world), as well as other problems which we turn to below.

It was precisely this emergence of classicality (as actually observed in our universe: the quasi-classical domain of ordinary experience) that concerned Gell-Mann and Hartle, with their histories amounting to what they called ``realms,'' defined by their mutual isolation from quantum interference (i.e. they are decoherent,  enabling systems to exhibit apparently definite properties). Since there was no observer required, only processes endogenous to quantum mechanics, the emergence of a classical universe could be modelled using only quantum mechanics, with no additional collapse postulates or hidden variables. One simply has a theory that specifies probabilities for alternative consistent histories for the universe, with the decoherence condition allowing probabilities to be thus assigned. Hartle refers to this framework, with its merging of the ideas of consistent histories, decoherence theory and cosmology, as ``Post-Everett'' quantum mechanics (see \citet[5-6]{Hartle1992} and \cite{Hartle1991}).
 




 Prior to his work with Gell-Mann, Hartle had  been working with Stephen Hawking applying path-integral techniques to Hawking's only just discovered theory that black holes emit radiation. The difficulty they faced in this paper was applying the (flat-space) quantum field theoretic technique of path-integration in the context of a curved spacetime background, in which a black hole appears. This led to their classic paper on the wavefunction of the universe, in 1983. The basis of this work is the Wheeler-DeWitt equation:

\begin{equation}
\mathcal{H}\Psi[h_{ij}, \phi] = 0	
\end{equation}

\noindent This is solved by appropriate universal wavefunctions, $\Psi[h_{ij}, \phi]$, of the 3-metric $h_{ij}$ on a 3-surface and the configuration of matter on that 3-surface. This is of course an Hamiltonian approach (the `$\mathcal{H}$' denotes the Hamiltonian): one develops the wave function from an initial time-slice to other times. The Wheeler-DeWitt equation is essentially just a kind of Schr\"{o}dinger equation, describing the time-development of wave-functions, though here these functions depend on the metric on a hypersurface and the embedding of that hypersurface into spacetime, rather than position and time. This equation develops the wave function deterministically, with only the various components of the superposition having amplitudes assigned.

 The decoherent histories paper, with Gell-Mann, needs to be understood against this quantum cosmological work, which was an attempt to make quantum mechanics apply to the universe as a whole.  If it is indeed the case that quantum mechanics is not limited in its application, then there is no reason why the universe should not also be described by a wavefunction. But this then stands in need of an interpretation, and clearly the Copenhagen interpretation falls short here: what serves as the classical domain when the universe itself is subject to quantization? What counts as an external observer in this case, where, as Bryce DeWitt puts it, ``the external observer has no place to stand'' \citep[319]{DeWitt1968}?

However, the idea here is that quantum mechanics now assigns probabilities to possible, consistent histories of the universe rather than measurement outcomes by classical observers. As mentioned, Gell-Mann and Hartle applied the notion of decoherence  to ground the classical probabilities of the consistent histories approach, hence their name ``decoherent histories'' to describe those histories that can represent classical universes of the kind we observe (i.e. with no strange interfering between histories). The central object describing the non-interfering alternative histories (as well, of course, as the degree of interference) is the ``decoherence functional'' which involves pairs of such (coarse-grained) alternatives as given in a sum-over-histories formulation (and so spits out a complex number when fed such pairs):

\begin{equation}
	D(\alpha' , \alpha) = N \int_{c_{\alpha'}}  \delta x' \int_{\alpha} \delta x \rho_{f}(x_{f} , x'{f})  \mathrm{exp}\{ i S[x'(\tau)] - S[x(\tau)])/\hbar \} \rho_{i}(x'_{0} , x_{0})  
\end{equation}

\noindent Here, the $c_{\alpha}$ are the exclusive classes of histories. This then grounds the probability sum rules of quantum mechanics since, a with decoherence as ordinarily understood (e.g. in the sense of Zeh), the off-diagonal terms of the decoherence functional represent the amount of interference between the alternative histories. When this interference is absent (or zero for all practical purposes), the diagonal terms provide the probabilities, which can be given a classical interpretation because of the negligible interference between them. 

Crucially, the selection of non-interfering alternatives is achieved not by some reduction process, resulting from measurement say, but by the specification of an appropriate initial condition and Hamiltonian. This is seemingly required in situations involving genuinely closed quantum systems. That is, in the case of closed system, decoherence replaces measurement as the deciding factor in the emergence of a classical world, as it must if both observer and observed are contained in the system.   But interpretational issues  are not entirely side-stepped here simply because measurement and observation have been dethroned.

Gell-Mann and Hartle appear to view both Copenhagen and Everett as approximations in some sense of their more fundamental interpretation of closed systems. A quasi-classical world emerges to provide a home for Copenhagen interpretations, and if one restricts the analysis of probabilities for correlations to a single moment of time (a Now), then Everettian interpretations are also provided for (see, e.g., \citeauthor{Hartle1991}, \citeyear[154]{Hartle1991}). Indeed, in acknowledgement of the ability to view decoherent histories as a kind of generalization of many worlds, Hartle proposes the title ``many histories'' for his own approach \cite[p. 155]{Hartle1991}.  

But it is somewhat difficult to nail down exactly what the interpretation is in the case of decoherent histores. Certainly, Griffiths believes that he presents a realist interpretation, on account of the absence of observers. But to do this, it seems as though the reality of many histories must be included, otherwise, on measurement we face a situation in which all but one of the histories are wiped out. In other words, on these histories approaches the probabilities for the histories are not probabilities for `getting' such a history on a measurement of a sequence of events, but probabilities \emph{for} those properties themselves, irrespective of whether or not a measurement is made (that is, they are beables). But then, it seems, we face a problem of predictions, since there are no outcomes here only records \emph{qua} correlations which appear classical thanks to decoherence, as in modern Everettian interpretations. In which case, realism about the consistent histories just is many worlds. When discussing the issue of ``reality,'' Gell-Mann and Hartle tend to allow only what is internal to the formalism to have any say. In particular, decoherence and the approximate emergence of classicality that it generates (for some suitable initial conditions) is all that matters.\footnote{Note that an observer (even as modelled endogenously) is not required for the emergence of a quasi-classical realm. Gell-Mann and Hartle speak more generally of ``measurement situations.'' There are certain operators that ``habitually decohere'' as a result, thus generating the quasi-classical realm (see \citeauthor{Gell_Mann-Hartle1994a}, \citeyear{Gell_Mann-Hartle1994a}). However, it must still be admitted that any coarse-graining (which generates some level of complexity) is a matter of selection.} Given this internalist viewpoint, all that quantum mechanics can do, on the question of which alternative history is `more real,' is supply a probability. This naturally sends the position into many-worlds-type interpretive issues, since one can then ask about the status of the ensemble of all alternatives, the history space (by analogy with the multiverse). 

The problem is that we do have a clear experience of one history happening.  Gell-Mann and Hartle bite the many worlds bullet, but add in a story about why we experience our specific universal history, again with a story that is supposed to be internal to quantum mechanics itself, employing Everett's innovation of including the observer as part of the overall quantum mechanical story (see \citeauthor{Gell_Mann-Hartle1994a},\citeyear[325]{Gell_Mann-Hartle1994a}). Fay Dowker (a PhD student of Hartle's) and Adrian Kent argued that the decoherent histories programme is really a promissory note requiring that ``a to-be-found theory of experience will identify as the fundamentally correct ones among all the possibilities offered by the formalism'' \citep[1641]{Dowker-Kent1996}. They suggest that quantum mechanics cannot supply everything, and must be supplemented by a ``selection principle'' (added as an additional axiom) to adequately capture our experience. \label{intprob} The reason is natural selection: while there are many co-existing histories each giving a `way the world could be,' we have evolved in such a way to pick out ours. This addition bears the hallmarks of the Santa Fe Institute, with its focus on the role of complexity in the universe. In this case, complexity is fundamental in an almost Kantian manner: without complexity there would be no organisms capable of experiencing the specific history we do in fact observe. However, it is not so very far from Griffiths own position on the status of the histories, which is that we choose for our convenience, or even \emph{aesthetic} reasons (see his comments after Bryce DeWitt's talk at the 1994 Mazagon conference: \citeauthor{DeWitt1994}, \citeyear[231]{DeWitt1994})! The concept of an ``IGUS'' (or information gathering and utilizing system) grounds the explanation for Gell-Mann and Hartle. Our sense of classical, single-historied reality amounts to a pattern of agreement between a kind of ecosystem of IGUSes (that is, \emph{us}, as modelled in physics) on the quasi-classical realm, which then seems robust and invariant (i.e. real). The IGUSes come with coarse-grainings of the world appropriate to their constitution (as localised entities with limited sensory capabilities, and so on, \cite[121]{Hartle1993}. They are also equipped with a memory, which involves correlations between the projection operators of the quasi-classical realm and of their internal-states (e.g. brain-states). This near-isomorphic mapping between projection operators is what constitutes experience or perception of a highly classical world. Hence, the concept of reality is somewhat contorted here: it is not a `world out there' (which, it seems, \emph{must} be some kind of history-ensemble). Instead, it is a relational construct involving correlations between memory records (themselves contingent on the initial condition for the universe along with the dynamics) and a specific history that is then naturally selected. In other words: this is an anthropic notion, though not one that treats the anthropic components independently of the other basic physical principles: it is more of a bootstrapped conception reality.

In this way, the separate elements of Copenhagen approaches are dealt with endogenously. It is the \emph{particular} coarse-grainings that the IGUSes settle upon that is to be described by evolution.\footnote{See also \cite{saund,saund2} along these same lines.} The response to Zurek's puzzlement is this: while quantum mechanics itself does not show any preference for one alternative history or set of such over another (pace probabilities), the IGUSes (or observers) in the world \emph{do} show a preference, for only some decoherent sets of histories are suitable from the point of view of an IGUS. In this response, we find a curious blending of the complexity work of Gell-Mann's Santa Fe Institute since an IGUS just is a kind of  complex adaptive system concerned with its survival \citep{Gell_Mann-Hartle1994a} with the cosmological approach of Hartle.

 If anything, rather than using a classical world to get a handle on the quantum world, as in standard Copenhagen interpretations, here one generates an approximation of the Copenhagen's classical world through quantum mechanics itself (cf. \cite[179]{harty88}). In this case, no collapse and no interaction between the classical and quantum worlds are required, with the former essentially just \emph{being} the latter, only from a different point of view, and one that emerges only given certain special initial conditions and only then emerges fairly late in the universe's evolution. That is, whereas Copenhagen involves a dualism, the decoherent histories approach, like Everettian approaches, is monist. The separate postulation of a classical world is, in Hartle's words, ``excess baggage' caused by the initial focus on laboratory experiments \citep[6]{Hartle1992}. Of course, what remains is to explain the initial condition of the universe at the root of the emergence of the reality we observe, which allows for records of an apparently unique classical history. This is largely dependent on the production of a quantum theory of gravity.

Ultimately, the links between the relative-state interpretation and decoherent histories became more apparent, with the latter simply supplying a solution to one of the key problems with the former (see, e.g., \cite{wal2003}). In the many-worlds approach of David Wallace and Simon Saunders, the decoherent histories function to select those classical worlds that follow probability theory with the various alternatives summing to unity (this unitarity is of course just what it means to be consistent here). In other words, decoherence of a family of histories implies that they can be treated as separate classical, exclusive alternatives each with an associated probability. Hartle himself later referred to his approach with Gell-Mann as an ``extension and to some extent a completion of Everett's work'' \cite[p. 986]{Hartle2011}.


\section{Conclusion}

Common to all history-based interpretations of quantum mechanics is the notion that one should be able to extract from it (rather than put in by hand) the everyday classical world. The basic idea is that one can treat families of histories using probabilities more or less along the lines of classical stochastic theory, with quantum mechanics then assigning probabilities to each possible history in such a family, so long as certain consistency conditions are satisfied. While the original proposals of \cite{griffiths1984} and \cite{Omnes1987} utilised predominantly formal consistency conditions, the later approach of \citet[ - which Griffiths would later also adopt]{Gellmann-Hartle1990} used decoherence to ground the proper assignment of probabilities to histories, which are then capable of describing the kind of world we see. There is also no special meaning assigned to measurement and observation (and, indeed observers) in such interpretations: these simply constitute just another process to be modeled within the formalism (we showed how cybernetic and complexity ideas provided a hospitable research landscape for modeling observers and their observations). This was the main motivation behind Gell-Mann and Hartle's approach since they desired an interpretation fit for cosmological applications, in which external measurement and observers make no sense. This link to quantum cosmology (and quantum gravity) has certainly led to the increased endurance of the consistent histories approach, and new developments and applications continue to appear as a result.

\bibliographystyle{apalike}
\bibliography{CHRefs}

\begin{thebibliography}{}

\bibitem[Barret and Byrne, 2012]{Barret-Byrne2012}
Barret, J.~A. and Byrne, P., editors (2012).
\newblock {\em The Everett Interpretation of Quantum Mechanics: Collected Works
  1955-1980 with Commentary}.
\newblock Princeton University Press.

\bibitem[Bell, 1990]{bell1990}
Bell, J.~S. (1990).
\newblock {Against {`measurement'}}.
\newblock In Miller, A.~I., editor, {\em Sixty-Two Years of Uncertainty.
  Historical, Philosophical, and Physical Inquiries into the Foundations of
  Quantum Mechanics}, pages 17--32. New York: Plenum Press.

\bibitem[Burtt, 1924]{Burtt1924}
Burtt, E.~A. (2003 [1924]).
\newblock {\em The Metaphysical Foundations of Modern Physical Science}.
\newblock Dover Publications Inc.

\bibitem[Cowan, 2010]{Cowan2010}
Cowan, G. (2010).
\newblock {\em Manhattan Project to the Santa Fe Institute: The Memoirs of
  George A Cowan}.
\newblock University of New Mexico Press.

\bibitem[Daston and Galison, 2007]{DastonGalison2007}
Daston, L. and Galison, P. (2007).
\newblock {\em Objectivity}.
\newblock Zone Books, Brooklyn, NY.

\bibitem[DeWitt, 1968]{DeWitt1968}
DeWitt, B. (1968).
\newblock The everett-wheeler interpretation of quantum mechanics.
\newblock In DeWitt and Wheeler, J., editors, {\em Batelle Rencontres}, page
  318–332. W. A. Benjamin, Inc.

\bibitem[DeWitt, 1994]{DeWitt1994}
DeWitt, B. (1994).
\newblock Decoherence without complexity and without an arrow of time.
\newblock In {\em Physical Origins of Time Asymmetry}, page 221–233.
  Cambridge University Press.

\bibitem[Dowker and Kent, 1996]{Dowker-Kent1996}
Dowker, F. and Kent, A. (1996).
\newblock On the consistent histories approach to quantum mechanics.
\newblock {\em Journal of Statistical Physics}, 82(5/6):1575–1646.

\bibitem[Duhem, 1906]{Duhem1906}
Duhem, P. M.~M. (1991 [1906]).
\newblock {\em The Aim and Structure of Physical Theory}.
\newblock Princeton University Press, Princeton.

\bibitem[Dupuy, 1994]{Dupy1994}
Dupuy, J.-P. (1994).
\newblock {\em On the Origins of Cognitive Science: The Mechanization of the
  Mind}.
\newblock The MIT Press.

\bibitem[Foucault, 1969]{Foucault1969}
Foucault, M. (1969).
\newblock {\em The Archaeology of Knowledge and the Discourse on Language}.
\newblock Pantheon Books.

\bibitem[Freire, 2015]{Freire2015}
Freire, O. (2015).
\newblock {\em The Quantum Dissidents: Rebuilding the Foundations of Quantum
  Mechanics (1950-1990)}.
\newblock Springer.

\bibitem[Gell-Mann, 1990]{Gell_Mann1990}
Gell-Mann, M. (1990).
\newblock {\em The Santa Fe Institute}.
\newblock Santa Fe Institute, Santa Fe, New Mexico.

\bibitem[Gell-Mann and Hartle, 1990a]{Gellmann-Hartle1990}
Gell-Mann, M. and Hartle, J. (1990a).
\newblock Alternative decohering histories in quantum mechanics.
\newblock In Phua, K. and Yamaguchi, Y., editors, {\em Proceedings of the 25th
  International Conference on High Energy Physics}. World Scientiﬁc,
  Singapore.

\bibitem[Gell-Mann and Hartle, 1990b]{Gell_Mann-Hartle1990b}
Gell-Mann, M. and Hartle, J.~B. (1990b).
\newblock {Quantum mechanics in the light of quantum cosmology}.
\newblock In Zurek, W.~H., editor, {\em Complexity, entropy and the physics of
  information}, pages 321--343. Reading: Addison Wesley, Boston.

\bibitem[Gell-Mann and Hartle, 1994a]{Gell_Mann-Hartle1994a}
Gell-Mann, M. and Hartle, J.~B. (1994a).
\newblock Strong decoherence.
\newblock In Feng, D.~H. and Hu, B., editors, {\em Quantum Classical
  Correspondence: Proceedings of the 4th Drexel Symposium on Quantum
  Nonintegrability}, pages 3--35. International Press.

\bibitem[Gell-Mann and Hartle, 1994b]{Gell_Mann-Hartle1994b}
Gell-Mann, M. and Hartle, J.~B. (1994b).
\newblock Time symmetry and asymmetry in quantum mechanics and quantum
  cosmology.
\newblock In Halliwell, J.~J., P\'{e}rez-Maercader, and Zurek, W., editors,
  {\em Physical Origins of Time Asymmetry}, pages 311--345. Cambridge
  University Press.

\bibitem[Gell-Mann and Hartle, 2007]{Gell_Mann-Hartle2007}
Gell-Mann, M. and Hartle, J.~B. (2007).
\newblock Quasiclassical coarse graining and thermodynamic entropy.
\newblock {\em Physical Review A}, 76(2):1--16.

\bibitem[Griffiths, 1984]{griffiths1984}
Griffiths, R.~B. (1984).
\newblock Consistent histories and the interpretation of quantum mechanics.
\newblock {\em Journal of Statistical Physics}, 36(1-2):219--272.

\bibitem[Griffiths, 1986]{griffiths1986}
Griffiths, R.~B. (1986).
\newblock Making consistent inferences from quantum measurements.
\newblock In Greenberger, D., editor, {\em New Techniques and Ideas in Quantum
  Measurement Theory}, pages 512--7. New York Academy of Sciences.

\bibitem[Griffiths, 1994]{griffiths1994}
Griffiths, R.~B. (1994).
\newblock A consistent history approach to the logic of quantum mechanics.
\newblock In {\em Symposium on the Foundations of Modern Physics 1994: 70 Years
  of Matter Waves, Helsinki, Finland, 13-16 June 1994}, pages 85--98. Editions
  Fronti{\`e}res.

\bibitem[Griffiths, 1996]{griffiths1996}
Griffiths, R.~B. (1996).
\newblock Consistent histories and quantum reasoning.
\newblock {\em Physical Review A}, 54(4):2759.

\bibitem[Griffiths, 2003]{griffiths2003}
Griffiths, R.~B. (2003).
\newblock {\em Consistent Quantum Theory}.
\newblock Cambridge University Press.

\bibitem[Griffiths, 2009]{Griffiths2009}
Griffiths, R.~B. (2009).
\newblock Consistent histories.
\newblock In Greenberger, D. e.~a., editor, {\em Compendium of Quantum Physics:
  Concepts, Experiments, History and Philosophy}, pages 117--122. Springer.

\bibitem[Hacking, 1990]{Hacking1990}
Hacking, I. (1990).
\newblock {\em The Taming of Change}.
\newblock Cambridge University Press.

\bibitem[Halliwell, 1989]{halliwell1989}
Halliwell, J.~J. (1989).
\newblock Decoherence in quantum cosmology.
\newblock {\em Physical Review D}, 39(10):2912.

\bibitem[Hartle, 1989]{harty88}
Hartle, J. (1989).
\newblock Quantum cosmology and quantum mechanics.
\newblock In Kafatos, M., editor, {\em Bell's Theorem, Quantum Theory and
  Conceptions of the Universe}, pages 179--180. Kluwer Academic.

\bibitem[Hartle, 1991]{Hartle1991}
Hartle, J. (1991).
\newblock The quantum mechanics of cosmology.
\newblock In Coleman, S., Hartle, J.~B., Piran, T., and Weinberg, S., editors,
  {\em Quantum Cosmology and Baby Universes}, pages 65--158. World Scientific.

\bibitem[Hartle, 1992]{Hartle1992}
Hartle, J. (1992).
\newblock Excess baggage.
\newblock In Schwarz, J., editor, {\em Elementary Particles and the Universe:
  Essays in Honor of Murray Gell-Mann}, pages 1--16. Cambridge University
  Press.

\bibitem[Hartle, 1993]{Hartle1993}
Hartle, J. (1993).
\newblock The quantum mechanics of closed systems.
\newblock In Hu, B.~L., Jr, M. P.~R., and Vishveshwara, C.~V., editors, {\em
  Directions in General Relativity: Volume 1: Proceedings of the 1993
  International Symposium, Maryland: Papers in Honor of Charles Misner}, pages
  104--124. Cambridge University Press.

\bibitem[Hartle, 1995]{hartle1995}
Hartle, J.~B. (1995).
\newblock Spacetime quantum mechanics and the quantum mechanics of spacetime.
\newblock In Julia, B. and Zinn-Justin, J., editors, {\em Gravitation and
  Quantizations: Proceedings of the 1992 Les Houches Summer School}, pages
  285--480. Elsevier.

\bibitem[Hartle, 2007]{Hartle2007}
Hartle, J.~B. (2007).
\newblock Quantum physics and human language.
\newblock {\em Journal of Physics A: Mathematical and Theoretical},
  40(12):3101--3121.

\bibitem[Hartle, 2011]{Hartle2011}
Hartle, J.~B. (2011).
\newblock The quasiclassical realms of this quantum universe.
\newblock {\em Foundations of physics}, 41(6):982--1006.

\bibitem[Heims, 1991]{Heims1991}
Heims, S.~J. (1991).
\newblock {\em The Cybernetics Group}.
\newblock The MIT Press.

\bibitem[Hohenberg, 2010]{hohenberg2010}
Hohenberg, P.~C. (2010).
\newblock Colloquium: An introduction to consistent quantum theory.
\newblock {\em Reviews of Modern Physics}, 82(4):2835.

\bibitem[Isham, 1994]{Isham1}
Isham, C.~J. (1994).
\newblock Quantum logic and the histories approach to quantum theory.
\newblock {\em Journal of Mathematical Physics}, 35(5):2157--2185.

\bibitem[Isham and Linden, 1994]{Isham2}
Isham, C.~J. and Linden, N. (1994).
\newblock Quantum temporal logic and decoherence functionals in the histories
  approach to generalized quantum theory.
\newblock {\em Journal of Mathematical Physics}, 35(10):5452--5476.

\bibitem[Joos et~al., 2003]{jooszehkiefer2003}
Joos, E., Zeh, H., Kiefer, C., Giulini, D., Kupsch, J., and Stamatescu, I.-O.
  (2003).
\newblock {\em Decoherence and the Appearance of a Classical World in Quantum
  Theory}.
\newblock Berlin, Heidelberg: Springer, second edition.

\bibitem[Joos and Zeh, 1985]{JoosZeh1985}
Joos, E. and Zeh, H.~D. (1985).
\newblock The emergence of classical properties through interaction with the
  environment.
\newblock {\em Zeitschrift f\"{u}r Physik B Condensed Matter}, 59(2):223--243.

\bibitem[Lafontaine, 2004]{Lafontaine2004}
Lafontaine, C. (2004).
\newblock {\em O Império Cibernético: Das Máquinas de Pensar ao Pensamento
  Máquina}.
\newblock Instituto Piaget Divisão Editorial.

\bibitem[Lovejoy, 1936]{Lovejoy1936}
Lovejoy, A.~O. (1936).
\newblock {\em The Great Chain of Being: A Study of the History of an Idea}.
\newblock Harvard University Press.

\bibitem[Meyerson, 1908]{Meyerson1098}
Meyerson, E. (1908).
\newblock {\em Identity and Reality}.
\newblock George Allen \& Unwin Ltd, London, 3rd edition.

\bibitem[Omnès, 1987]{Omnes1987}
Omnès, R. (1987).
\newblock Interpretation of quantum mechanics.
\newblock {\em Physics Letters A}, 125(4):169--172.

\bibitem[Omnès, 1988a]{Omnes1988a}
Omnès, R. (1988a).
\newblock {Logical reformulation of quantum mechanics. I. Foundations}.
\newblock {\em Journal of Statistical Physics}, 53(3--4):893–--932.

\bibitem[Omnès, 1988b]{Omnes1988b}
Omnès, R. (1988b).
\newblock {Logical reformulation of quantum mechanics. II. Interferences and
  the Einstein-Podolsky-Rosen experiment}.
\newblock {\em Journal of Statistical Physics}, 53(3-4):933–955.

\bibitem[Omnès, 1988c]{Omnes1988c}
Omnès, R. (1988c).
\newblock {Logical reformulation of quantum mechanics. III. Classical limit and
  irreversibility}.
\newblock {\em Journal of Statistical Physics}, 53(3-4):957–975.

\bibitem[Omnès, 1989]{Omnes1989}
Omnès, R. (1989).
\newblock {Logical reformulation of quantum mechanics. IV. Projectors in
  semiclassical physics}.
\newblock {\em Journal of Statistical Physics}, 57(1-2):357–382.

\bibitem[Omnès, 1994]{Omnes1994}
Omnès, R. (1994).
\newblock {\em The Interpretation of Quantum Mechanics}.
\newblock Princeton University Press, Princeton, NJ.

\bibitem[Omnès, 1999]{Omnes1999}
Omnès, R. (1999).
\newblock {\em Understanding Quantum Mechanics}.
\newblock Princeton University Press, Princeton.

\bibitem[Omnès, 2005]{Omnes2005}
Omnès, R. (2005).
\newblock {\em Converging Realities: Toward a Common Philosophy of Physics and
  Mathematics}.
\newblock Princeton University Press, Princeton.

\bibitem[Omnès and Sangalli, 1999]{Omnes-Sangalli1999}
Omnès, R. and Sangalli, A. (1999).
\newblock {\em Quantum Philosophy: Understanding and Interpreting Contemporary
  Science}.
\newblock Princeton University Press, Princeton.

\bibitem[Pines, 1988]{Pines1988}
Pines, D. (1988).
\newblock {Designing a University for the Millennium: A Santa Fe Institute
  perspective}.
\newblock In {\em In A keynote address to the April 1998 Fred Emery Conference
  of Sabanci University}, Istanbul, Turkey.

\bibitem[Ricoeur, 1963]{Ricoeur1963}
Ricoeur, P. (1963).
\newblock {\em L’Herme´neutique et le Structuralisme}.
\newblock L, Esprit: Novembre.

\bibitem[Saunders, 1993a]{saund2}
Saunders, S. (1993a).
\newblock Decoherence and evolutionary adaptation.
\newblock {\em Physics Letters A}, 184(1):1--5.

\bibitem[Saunders, 1993b]{saund}
Saunders, S. (1993b).
\newblock Decoherence, relative states, and evolutionary adaptation.
\newblock {\em Foundations of physics}, 23(12):1553--1585.

\bibitem[Wallace, 2003]{wal2003}
Wallace, D. (2003).
\newblock Everett and structure.
\newblock {\em Studies in History and Philosophy of Science Part B: Studies in
  History and Philosophy of Modern Physics}, 34(1):87--105.

\bibitem[Zeh, 1970]{zeh1970}
Zeh, H.~D. (1970).
\newblock On the interpretation of measurement in quantum theory.
\newblock {\em Foundations of Physics}, 1(1):69--76.

\bibitem[Zurek, 1981]{Zurek1981}
Zurek, W.~H. (1981).
\newblock Pointer basis of quantum apparatus: Into what mixture does the wave
  packet collapse?
\newblock {\em Physical Review D}, 24(6):1516--–1525.

\end{thebibliography}
\end{document}